\def\n{\noindent}

\def\b{\bigskip}

\def\kappaplus{(\kappa + \sigma)}
\def\kappaminus{(\kappa - \sigma)}

\def\F{{\bf F}}
\def\u{{\bf u}}
\def\ut{\frac{\partial {\bf u}}{\partial t}}
\def\rhot{\frac{\partial \rho}{\partial t}}
\def\Et{\frac{\partial E}{\partial t}}
\def\Ft{\frac{\partial {\bf F}}{\partial t}}
\def\dz{\frac{\partial}{\partial z}}

\def\sub#1{_{_{#1}}}

\def\hexnumber#1{\ifcase#1 0\or1\or2\or3\or4\or5\or6\or7\or8\or9\or
  A\or B\or C\or D\or E\or F\fi }

\font\tenbms=cmbsy10
\font\sevenbms=cmbsy10 at 7pt
\font\fivebms=cmbsy10 at 5pt

\newfam\bmsfam
\textfont\bmsfam=\tenbms
\scriptfont\bmsfam=\sevenbms
\scriptscriptfont\bmsfam=\fivebms

\edef\bsy@{\hexnumber\bmsfam}

\mathchardef\bnabla="0\bsy@72 
\documentstyle{article}
\begin{document}

\title{PHOTOFLUID INSTABILITIES OF HOT STELLAR ENVELOPES}
\author{Edward A. Spiegel and Louis Tao\\
{\it Department of Astronomy} \\ {\it Columbia University} \\
{\it New York, NY 10027} }
\maketitle

\begin{abstract}
Beginning from a relatively simple set of dynamical equations for
a fluid permeated by a radiative field strong enough to produce
significant forces, we find the structure of plane-parallel equilibria
and study their stability to small acoustic disturbances.  In doing this,
we neglect viscous effects and complications of nongreyness.  We find
that acoustic instabilities occur over a wide range of conditions
below the Eddington limit.  This result is in line with findings reported
twenty years ago but it contradicts some more recent reports of the
absence of instabilities.  We briefly attempt to identify the causes of
the discrepancies and then close with a discussion of the possible
astrophysical interest of such instabilities.
\end{abstract}

\n
\leftline{PACS: 47.70.Mc, 97.10Ex}
\leftline{Keywords: radiation gas dynamics, stellar atmospheres, stellar
activity and pulsation}

\section{Introduction}
\renewcommand{\theequation}{1.\arabic{equation}}
\setcounter{equation}{0}
Evidence of high levels of fluid dynamical activity in hot stellar
atmospheres has been available since O. Struve detected large line
widths that he attributed to macroturbulence.  In fact, the actual
nature of the motions is still not certain.  Given the many possible
causes of vigorous motion, this is not surprising, and it remains unclear
whether the activity is driven by rotation, pulsation, radiation,
or all of these.  Huang and Struve (1960) favored the notion that the large
partial pressure of radiation in hot stars plays a key r\^ole in this
problem, but they never made it clear how.  Indeed, in the context of
early cosmology, radiation pressure is thought of as a stabilizing
influence.

In this discussion we wish to isolate the effects of the radiative forces
and to discuss the possible consequences they may have for the fluid
dynamics of a stratified layer.  It goes without saying that aspects of
this topic should be of interest for stellar interiors and hot disks as
well as stellar atmospheres.

A concrete image of the kind of fluid motions that radiative forces
might produce is to be found in the dynamics of fluidized beds
(Davidson and Harrison, 1963).  Fluidization occurs when a fluid is
forced to flow upward through a bed of particles. When the upward drag
on each particle balances its weight, the particles are levitated
and they form a second fluid.  The bed is then said to be fluidized and,
once this happens, it is in a state analogous to that of a stellar
atmosphere at the Eddington limit, when radiative levitation balances
gravity (Prendergast and Spiegel, 1973).

In fluidized beds, the drag per particle is a sensitive function
of the density of particles.  Hence, the fluidized state can be sustained
over a wide gamut of conditions ranging from the lethargic case of
quicksand to the kind of vigorous bubbling that may also
occur in a hot stellar atmosphere close to the Eddington limit.
The bubbling state arises when the typical mass density of a single
particle in the bed greatly exceeds the density of the driving fluid.  In
that case, voids in the particle distribution form, rise to the top of
the bed and collapse, throwing particles upward, as in a boiling liquid.
This phenomenon has interested chemical engineers for some time and, if
the analogy to the astrophysical situation holds, there is much interest
in it for the astrophysicist as well.

While it is not clear exactly how fluidization bubbles form, it is
generally supposed that the formation process is driven by instability.
In the case of the fluidized bed, there are known instabilities resulting
from the dependence of the drag per particle on the density of particles.
This mechanism is thus a mechanical analogue of the $\kappa$-mechanism of
stellar pulsation.
Under suitable conditions, when such
mechanisms operate, we can expect a Hopf bifurcation that, in the
stellar case, causes pulsation.

The hottest stars are mostly ionized and so they are not highly
susceptible to instability by the
$\kappa$-mechanism that operates in many cooler stars.  Nevertheless,
very hot stars do pulsate.  That may be sufficient grounds for
suspecting that bubbling occurs in their atmospheres, since pulsation can
produce parametric instability of convective modes in a star (Poyet and
Spiegel, 1979).  In very hot stars, pulsation may be driven from the
stellar cores or be the result of instability in the outer layers
(Umurhan, 1998). In the
present work we investigate the possibility of radiatively driven
instabilities in hot stellar atmospheres, or rather, in hot slabs
stratified under gravity and radiative forces.

The stability problem of hot stellar atmospheres has been studied for
a few decades with mixed results.  Over twenty years ago, many
participants at a meeting in Nice (Cayrel, R. and Steinberg, M. 1976)
concluded that there were instabilities in very hot atmospheres.   At that
time, the main issue under discussion was whether the instabilities were
convective or absolute.  However, subsequent studies offered the
conclusion that radiatively driven instability could occur only under very
special conditions if at all.  Marzek (1977) found that instability
occurred only just below the Eddington limit.  Marzek worked directly with
the transfer equation by numerical means and thus had more detail in his
system than most previous investigators.  Recently, Asplund (1998)
claimed that there is no instability until the Eddington limit is
surpassed.  In his treatment of the transfer theory, Asplund omitted
the same terms that many early workers dropped in the closure of the
transfer equations and so had less detail than some previous workers.
So the different conclusions of the older and newer
work is not just a question of more or less detail in the basic equations.

The reason for the change in the prevailing results is not clear.  It
may just be that workers who had reproduced the earlier results simply
saw no point in announcing this, so they have not come forward.  The
results of Marzek are also unpublished.
On the other hand, the observed evidence of great activity in the
hottest stars does suggest some form of instability.  Perhaps, there
are other mechanisms at work.  In this vein, Arons (1992) has included
magnetic effects to promote instability. There are also no doubt
rotational instabilities and parametric instabilities that may be present.
However, the issue of the radiative instabilities has become clouded and
it seems a sufficiently important astrophysical issue that the problem
should be reconsidered. That is the purpose of this paper.

In the next section we present the equations we shall use with some
discussion of their provenance.  Then we turn to the equilibria
they allow and linearize the full equations about them.  We do find
unstable acoustic waves, and make an
attempt to understand their origin before concluding with a brief
discussion of their possible implications.
\bigskip

\section{The Photofluid Equations}

\renewcommand{\theequation}{2.\arabic{equation}}
\setcounter{equation}{0}

For this study we use the continuum description of both the matter
and the radiation.  For each, we have a stress tensor and the vanishing
of the divergence of the sum of these provides the equations of the
problem, though they must be supplemented by equations of state for the
material and radiative fluids and by formulae for scattering and
absorption coefficients.

We express the equations in an inertial frame that
we call the basic frame.  In this case, the equations of motion are (Hsieh
and Spiegel, 1979; Simon 1963): \begin{eqnarray}
\rhot + \bnabla\cdot(\rho \u) &=& 0
\label{eq:continuity}\\
\rho\left[\ut + (\u \cdot \bnabla) \u\right] &=& -\bnabla p -
\rho g \widehat{\bf z}
+ \rho \frac{\kappaplus}{c} \left(\F -\frac{4}{3} E\u\right)
\label{eq:momentum}\\
\Et + \bnabla \cdot \F &=& \rho \kappa c (S - E) + \rho
\frac{\kappaminus}{c} \u \cdot \F
\label{eq:rad-E}\\
\Ft + \frac{c^2}{3} \bnabla E &=& -\rho \kappaplus c
(\F - \frac{4}{3} E\u) + \rho \kappa c (S-E)\u - \bnabla \cdot
(\u \F + \F \u) + \frac{2}{3} \bnabla (\u\cdot\F)
\label{eq:rad-F}\\
{\frac{\partial p}{\partial t}} + (\u \cdot \bnabla) p - c_s^2
\left[{\frac{\partial \rho}{\partial t}} + (\u \cdot \bnabla)\rho\right]
&=&
-(\gamma-1)\left[\rho \kappa c(S-E)
+ 2\rho \frac{\kappa}{c} \u\cdot\F\right]
\label{eq:entropy}\\
S &=& a T^4
\label{eq:LTE}\\
p &=& R \rho T
\label{eq:ideal}
\end{eqnarray}
In these equations, all quantities are expressed in the basic frame.
The fluid variables ${\bf u}, \rho, p$ and $T$ are the velocity,
the mass density, pressure and temperature, respectively.  The main
radiative variables, ${\bf F}$ and $E$ are the radiative flux, and energy
density, both integrated over frequency while $R$, $a$ and $\gamma$ are
the gas constant, the radiation constant and the ratio of specific heats.
The Thomson cross section is $\sigma$, $\kappa$ is the mean absorption
coefficient, $c_s$ is the speed of sound, and $c$ is the speed of light.

We adopt the plane-parallel geometry of an atmosphere with constant
gravity; the unit vector $\widehat{\bf z}$ points in the
vertical direction.  Since the radiation is relativistic, we have included
here some ${\cal O}(v/c)$ corrections to the radiative quantities.  For
example, the radiative flux as seen in a frame locally moving with the
matter is ${\bf F}-{4\over 3}E{\bf u}$.  Such corrections are not
quantitatively important in the present discussion, as we shall see.

In continuum physics one writes constitutive relations expressing the
stress in terms of other basic fluid properties.  Here we consider
an ideal fluid in which the dissipation is caused by interaction with a
coexisting radiation fluid.  For the latter,
a constitutive relation, or closure approximation, was derived from the
equation of transfer by taking its first two moments.  That system was
then closed by the Eddington relation between the radiative pressure
tensor and the radiative energy density, $P_{ij}={1 \over 3}E\delta_{ij}$
where $\delta_{ij}$ is the Kronecker delta.  The resulting system was then
expanded in ${\bf u}/c$, where $c$ is the speed of light, and truncated at
order ${\bf u}/c$.

A number of simplifications have been made in deriving the basic equations
given here.  We assumed that the energy exchanged between the fluid and
the radiation field may be qualitatively accounted for by absorption and
re\"emission.  Thomson scattering is assumed with the qualitative notion
that Compton effect can be represented as part of the absorption term.
The details of such processes have been blurred here because we have
replaced absorption coefficients by a single mean absorption coefficient.
Such replacement has been a standard astrophysical practice on which much
discussion has centered.  The best means to use depends on which term in
which equation one is focused on and we have simply assumed that they are
all the same.  The problem is even more delicate when the medium is in
motion, for then the means take on tensor character.  We have assumed that
these tensors are diagonal and that all their nonzero elements are equal.

We are working in the Eddington approximation, which represents the
radiative pressure as diagonal.  It therefore does not include viscous
stresses and we have not added any.  When those are included they are
likely to promote stability, unless they are anisotropic.  The effects
of bulk viscosity have also not been explored.  All these issues await
discussion in future work, but we feel the present level of complexity
makes a good starting point for revisiting the stability problem.

Finally, we need to worry about boundary conditions.  The ones we use
shall be mentioned in connection with the linear stability problem below.
These will be solved in a slab of finite thickness extracted from the
equilibrium structure, to which we turn next.

\section{An Equilibrium State}

\renewcommand{\theequation}{3.\arabic{equation}}
\setcounter{equation}{0}

The question of instability naturally involves a statement of
the state that is unstable.  Finding such states can in itself
be a complicated matter, especially in cases where the Eddington
limit has been exceeded.   Studies such as that of Asplund in
which the Eddington limit is exceeded over the whole layer are
in fact difficult to assess since it is not clear if such local
conditions can be matched to a realistic stellar model that has a
proper spherical shape both within and without.  In the books
of Eddington and of Chandrasekar on stellar structure one
finds some spherical models where the Eddington limit is locally
exceeded in the deep interior, but this reversal of the local
effective gravity is possible only if the opacity is suitably
dependent on physical conditions (Underhill, 1949).  Such
situations do not arise in an atmosphere dominated by Thomson
scattering.

We are here studying the plane parallel case and the only way
in which to have a portion of it exceed the Eddington limit in
a physically reasonable way is to have a suitably variable opacity.
This is not easily achieved in a very hot atmosphere where scattering
dominates so we shall not consider super-Eddington conditions here.
Indeed, we are here mainly interested in instabilities in more typical
stellar atmospheric conditions.

From equation (\ref{eq:entropy}) we find that in a steady state with
no motion, we must have $E=S=aT^4$.  Then we see that
${\bf F}=F \widehat{\bf z}$ with constant $F$.  For a state that is
horizontally homogeneous and depends only on the vertical coordinate, $z$,
with the flux in the $\widehat {\bf z}$-direction, the basic equations
are
the following.  The hydrostatic equation of the fluid is \begin{equation}
{d p \over dz} = g_* \rho \label{eq:hstat}
\end{equation}
where $p$ is the gas pressure and \begin{equation}
g_* = g - g\sub R \qquad {\rm with} \qquad
g\sub R = {\sigma + \kappa\over c} F
\ ; \label{eq:grav} \end{equation}
$g_*$ is the acceleration of gravity corrected for photolevitation.  The
analogous relation for the radiative pressure, ${1\over 3} E$ is
\begin{equation} {d E\over dz} = - 3 g\sub R \rho\ . \label{eq:hrad}
\end{equation} With these equations we also use the equation of state
(\ref{eq:ideal}).

This is a standard problem in the theory of atmospheres and by analogy
with what is done there we introduce an optical coordinate,
\begin{equation}
\tau = \int_0^z \rho (\sigma + \kappa) ds^\prime \ , \label{eq:optd}
\end{equation}
where the origin of $z$ is at some convenient level in the atmosphere
where we also locate the origin of $\tau$.  We see then that
\begin{equation}
E = \hat E - 3{F\over c} \tau\ , \label{eq:Esln} \end{equation}
where the integration constant, $\hat E$,
is the value of $E$ at $\tau = 0$.
This result is a standard of the Eddington approximation of radiative
transfer in a stellar atmosphere.

From equation (\ref{eq:hstat}) and the equation of state we readily
find the integral \begin{equation}
p = \hat p - gq + {F\over c} \tau \ , \label{eq:press} \end{equation}
where $\hat p$ is the pressure at the origin of $\tau$ and of $z$ and
\begin{equation}
q = \int_0^z \rho dz^\prime \label{eq:cden} \end{equation}
is a column density.  Thus, \begin{equation}
{\cal P} \equiv p + {1\over 3} E = \hat {\cal P} - gq \label{eq:totp}
\end{equation}
which indicates that the total pressure supports the weight of the
atmosphere.

At this point, it becomes complicated to discuss the general case
and it is best to separate the case of the hot atmosphere with $\sigma >>
\kappa$  from the opposite case of the cool atmosphere, or even that of
the earth's atmosphere.  In the former case, since $\sigma$ is constant,
we can write a closed expression for the properties.  In the latter case,
if we may approximate $\kappa$ as a product of powers of density and
temperature we can obtain similar results by using matched expansions.  In
the cool case, we may omit the radiative forces.
Nevertheless, in both cases, the atmosphere splits into two parts, a lower
polytropic layer and an upper isothermal layer.  We illustrate with an
idealized case that we shall concentrate on here, with
$\sigma + \kappa$ taken constant.

For our illustrative static atmosphere, we have
$\tau = (\kappa + \sigma)q$ and so (\ref{eq:press}) and (\ref{eq:totp})
combine into \begin{equation}
p = \hat p + {g_*\over 3g\sub R}(E-\hat E) \ . \label{eq:pE} \end{equation}
As $z\rightarrow \infty$, the matter thins out and we require that the
gas pressure should vanish in that limit.  Nevertheless, radiation pours
out of the atmosphere, so the radiation pressure need not vanish.  If
we use a tilde to denote evaluation at $z=\infty$, (\ref{eq:pE}) becomes
\begin{equation}
\tilde E = \hat E - {3g\sub R \over g_*} \hat p \ . \label{eq:tilde}
\end{equation}
We obtain
\begin{equation}
p = {g_*\over 3g\sub R}(E - \tilde E) \ .
\label{eq:Ep}
\end{equation}
Then, if we write $\tilde E = a \tilde T^4$ and $\Theta=T/\tilde T$,
(\ref{eq:hrad}) becomes \begin{equation}
4{{\cal R}\tilde T\over g_*} {d\Theta \over dz} =
{\Theta^4 - 1\over \Theta^4}    \ . \label{eq:Theta} \end{equation}
This is readily solved and we find (Spiegel, 1977) \begin{equation}
-{g_* z\over 4{\cal R}\tilde T} = \Theta - {1 \over 2}\tan^{-1} \Theta
- {1 \over 2}{\rm cotanh}^{-1}\left(\Theta\right) \ .
\label{eq:prof} \end{equation}

The temperature profile for this atmosphere is linear for $z$ very
negative and constant for $z$ very positive.  So we have two rather
distinct regions: a polytrope below and an isothermal atmosphere above.
When we set $\tilde T = 0$, we find that $T$ is linear and we have
simply a  polytropic atmosphere whose upper boundary is at $z=0$.
To complete the solution, we note that (\ref{eq:Ep}) gives us $p$ in
terms of $E$, that is, $aT^4$ and that the ideal gas law is then used
to compute $\rho$.

It is interesting that composite structures like this one occur in
cool star stellar atmospheres as well as in the earth's atmosphere.
In 1899 Teisserenc de Bort reported his discovery of (what was then
called) the {\it isothermal region} in the high atmosphere of the earth.
This layer is now known as the stratosphere and the early attempts to
understand it did involve radiative transfer as well as other
complications that were thought essential (Humphreys, 1964).  It is
interesting that the present theory gives rise to a composite
structure resembling that of the troposphere and stratosphere and seems
to give a rough description of the combination of the photosphere and
chromosphere of the sun as well.  In our simplified version, the
lower layer is limited to a polytropic exponent of $4/3$, but this is
readily generalized and, in any case, is quite appropriate to our present
interest in hot stars.

We need to stress that a plane-parallel model are to be used over
sufficiently confined range of depths, even though the solution just
mentioned extends over an infinite range of depths.  We shall
study the stability of only a finite slab cut out of this solution
in the next section.

\section{Stability Theory}

\renewcommand{\theequation}{4.\arabic{equation}}
\setcounter{equation}{0}

\subsection{Linear Equations}

To describe small perturbations to the static solution
just described,  we indicate the value of a variable in
equilibrium by the subscript naught.  The linearized equations are
separable in space and time and horizontal and vertical coordinates so
that a typical variable, such as density, has the form \begin{equation}
\rho(x,y,z,t) = \rho_0(z) + \rho_1(z) \exp(ikx - i\omega t)
\label{eq:pert}  \end{equation}
where the perturbation eigenfunction has to be determined along with
the eigenvalue $\omega$.

When we put the forms (\ref{eq:pert}) for all the variables into the basic
equations and neglect quadratic and higher terms we obtain a set of
equations for the perturbation quantities.  It is easier to read the
content of this set if we suppress the subscript one from the
perturbation amplitudes.  When we do this, the equations are
\begin{eqnarray}
i\omega\frac{\rho}{\rho_0} &=& ik u + \dz w +
w \frac{1}{\rho_0} \dz\rho_0\\
-i\omega\rho_0 w &=& -\dz p -g_* \rho +\rho_0 \frac{\kappaplus}{c}
\left(F_z - \frac{4}{3} E_0 w\right)\\
-i\omega\rho_0 u &=& -ik p +\rho_0\frac{\kappaplus}{c}\left(
F_x - \frac{4}{3}E_0 u\right)\\
-i\omega E + ikF_x +\dz F_z &=& \rho_0 \kappa c (S - E) +
\rho_0\frac{\kappaminus}{c} F_0 w\\
-i\omega F_x + ik\frac{c^2}{3}E &=&-\rho_0 \kappaplus c
\left(F_x -\frac{4}{3}E_0 u\right)
-F_0 \dz u + \frac{2}{3} i k F_0 w\\
-i\omega F_z + \frac{c^2}{3}\dz E &=& -\rho_0 \kappaplus c
\left(F_z -\frac{4}{3}E_0 w\right) - \rho \kappaplus c F_0
-ik F_0 u -\frac{4}{3} F_0 \dz w\\
-i\omega p - \rho_0 g_* w + c_s^2\rho_0( i k u +\dz w) &=&
-(\gamma-1)\rho_0\kappa c (S-E)
- 2 (\gamma-1)\frac{\rho_0\kappa F_0}{c}w\\
S &=& 4 E_0 (\frac{p}{p_0} - \frac{\rho}{\rho_0})
\label{eq:perturbation}
\end{eqnarray}

\n where $F_z$ is the $z$-component of the perturbed flux, and so on.

\subsection{Two-point boundary--value eigenvalue problem}

In our solutions of the linear equations, we have found that the terms
coming from frame changes have very little effect.  For example,
the last two terms of equation (\ref{eq:rad-F}) come from making the
Eddington approximation in the frame co-moving with the fluid and then
Lorentz-transforming back into the star frame.  We find that when we solve
the equations without those terms and then evaluate those from the
solutions so obtained, they are indeed quite small.  On the basis of such
consistency checks, we omit them.  The full set of reduced equations
is exhibited in the next section.

In the problem simplified in this way, there are four first-order
equations in $z$ for four dependent variables.  The ones we work with
are the vertical component of the velocity field, $w$, the gas pressure,
$p$, the vertical component of the radiation flux, $F_z$, and the
radiative energy density, $E$.  Thus we have the following set of first
order equations:
\b


\begin{equation}
\dz w = \left(\frac{N^2}{g_*} - \frac{g_*}{c_s^2}\right) w
- i k u + i\frac{\omega}{\rho_0} \rho
\label{eq:uz}
\end{equation}

\begin{equation}
\dz p = \left[i\omega - \frac{4}{3}\frac{\kappaplus E_0}{c}\right]
\rho_0 w + \frac{\rho_0\kappaplus}{c} F_z - g_* \rho
\label{eq:p}
\end{equation}


\begin{equation}
\dz F_z =  -\left(\rho_0\kappa c - i\omega\right) E
+ \frac{\rho_0\kappaminus F_0}{c} w + 4 \frac{\kappa E_0 c}{c_s^2}p
-i k F_x - 4 \kappa E_0 c \rho
\label{eq:fz}
\end{equation}

\begin{equation}
\dz E = -\frac{3}{c^2}\left[\rho_0\kappaplus c - i\omega\right] F_z
+ 4\frac{\rho_0\kappaplus E_0}{c} w - 3\frac{\kappaplus F_0}{c} \rho
\label{eq:e}
\end{equation}

\b\b
\n where $g_*$ is the effective gravity and
$N$ is the buoyancy frequency given by

\begin{equation}
\frac{N^2}{g_*} =
-\left(\frac{1}{\rho_0} \dz \rho_0 - \frac{g_*}{c_s^2}\right)
\label{eq:Brunt}
\end{equation}

\b
Furthermore, the density, horizontal velocity, and
horizontal radiative flux perturbations satisfy

\begin{equation}
\rho = \frac{1}{c_s^2} p -
\left[i\omega - 4\frac{(\gamma-1)\kappa c E_0}{c_s^2}\right]^{-1}
\frac{2(\gamma-1)\rho_0\kappa F_0}{c^2_s c} w +
\left[i\omega - 4\frac{(\gamma-1)\kappa c E_0}{c_s^2}\right]^{-1}
\frac{(\gamma-1)\rho_0\kappa c}{c_s^2} E
\label{eq:rho_perturbation}
\end{equation}

\begin{equation}
u =
\left(1+\frac{4}{3}\frac{E_0}{\rho_0 c^2} -
i\frac{\omega}{\rho_0\kappaplus c}\right)^{-1}
\left[\frac{k}{\omega \rho_0}
\left(1-i\frac{\omega}{\rho_0\kappaplus c}\right) p
+\frac{k}{3\omega \rho_0} E\right]
\label{eq:ux}
\end{equation}

\begin{equation}
F_x =
\left(1+\frac{4}{3}\frac{E_0}{\rho_0 c^2} -
i\frac{\omega}{\rho_0\kappaplus c}\right)^{-1}
\left[
\frac{4}{3}\frac{k E_0}{\omega\rho_0} p
+ \frac{4}{9}\frac{k E_0}{\omega\rho_0}
\left(1 - i\frac{3}{4} \frac{\omega c}{\kappaplus E_0}\right) E
\right]\ .
\label{eq:fx}
\end{equation}

We shall solve these equations in a slab of finite
vertical extent in the domain $z_b \le z \le z_t$.  The choices of
the lower and upper boundaries have a significant effect on the stability
results since we may have a nearly pure polytrope, an almost isothermal
atmosphere, or a composite of the two.
Though the degree of instability does
change in response to variations in such choices, the instabilities do
seem to occur over a range of choices.  In all cases we truncate the
equilibrium so that the background density is finite everywhere in the
computational domain.  This prescription minimizes the influence of
singularities in the equations.

We assume that there is no mass flow through the boundaries and so set
\begin{equation}
w(z_b) = w(z_t) = 0
\label{eq:bc1}
\end{equation}
We also adopt standard Eddington boundary conditions on the boundaries
(Hsieh, 1977), namely
\begin{eqnarray}
E(z_b) + 2 F_z(z_b) & =&  0 \quad\quad(bottom)\\
E(z_t) - 2 F_z(z_t) & =&  0 \quad\quad(top)
\label{eq:bc2}
\end{eqnarray}
\n to  fix the incident radiation at the lower boundary and to
ensure that no radiation comes from the upper boundary.

\subsection{Numerical results}
We have solved the equations numerically in a variety of parameter
ranges.  The reliability of the program we used has been checked by
independent calculations.  In brief, the main result of the study is that
we find instabilities in a wide range of conditions.  We give here a brief
sampling of the results in a parameter range that is not extreme from
the standpoints of stiffness and other numerical considerations.

For a given horizontal wave number, $k$, the equations and boundary
conditions (\ref{eq:uz}) - (\ref{eq:bc2}) form a two-point boundary value
eigenvalue problem with the (complex) frequency of oscillation $\omega$ as
the eigenvalue.  We seek solutions in a sequence of
horizontal wavenumbers to trace particular branches of modes.  In the
case without radiative effects, this procedure correctly recovers
the adiabatic dispersion relations that are known analytically.
However, in the present case with radiative driving, we have only
numerical results to offer so far. To obtain these, we have used a
Newton-Raphson-Kantorovich relaxation scheme to find and track the
various branches of radiative fluid waves.  Here we describe the unstable
photo-acoustic modes, which are the most likely to produce significant
observable effects.

We find instability even for the rather mild case of $\alpha = g_*/g =
0.9$, and for a wide range of physical parameters as we raise the level
of radiative levitation by lowering $\alpha$.  The horizontal wave number
of the fastest growing mode in each case depends sensitively on the
truncation of the basic state.  We find a $k=0$ instability for basic
states that are mostly polytropic; however, in general, the inclusion of
the isothermal region reduces the growth rate of the $k=0$ instability, 
and may even move the most unstable modes to finite wave numbers.
Here we illustrate the kind of results obtained with the case
$\alpha = 0.7$.

In Figs.\ 1-3 we show the eigenvalues as a function of horizontal
wave number $k$ for $\alpha=0.7$.  In each case the upper panel
is the real part of $\omega$, the frequency, and the lower panel
shows (minus) the imaginary part, or growth rate.  Fig.\ 1 is for
a polytrope, Fig.\ 2 is from the approximately isothermal region of the
equilibrium only, and Fig.\ 3 is for a composite containing the transition
region at $z=0$.  The time of growth of unstable modes are tens or
hundreds of oscillation periods, depending on the details of the
structure.  In fact, this represents reasonably strong instability by
the standards of pulsation theory.

The mode with $k=0$ corresponds to radial oscillation.  It is
typically the most unstable one in the purely polytropic case.  As we
go to the composite case, the wave number of maximum growth goes to
a finite value.  Even if the main instability were radial, it would
probably trigger higher wavenumbers through secondary instability
if its amplitude became large enough.  We show the eigenfunctions
corresponding to the $k=0$ case of Fig.\ 3 in Fig.\ 4; the panels on
the right are the real parts and those on the left the imaginary parts.
The eigenfunctions do not show appreciable dependence on $k$, so
we may spare the reader further details of these.

\section{Discussion}

\renewcommand{\theequation}{5.\arabic{equation}}
\setcounter{equation}{0}

Having computed the stability calculation for a variety of parameters,
we then evaluated the terms individually and, in this way, we
learned which terms are important for the stability.  These results
should be useful elsewhere in developing approximation techniques for
such problems.  Here we show the equations with these terms omitted.
They are negligible in the problems studied and probably remain so for all
nonrelativistic cases.  The reduced equations are

\begin{eqnarray}
\rhot + \bnabla\cdot(\rho \u) &=& 0
\label{eq:continuit*y}\\
\rho\left[\ut + (\u \cdot \bnabla) \u\right] &=& -\bnabla p -
\rho g \widehat{\bf z}
+ \rho \frac{\kappaplus}{c} \F
\label{eq:momentu*m}\\
\Et + \bnabla \cdot \F &=& \rho \kappa c (S - E)
\label{eq:rad-*E}\\
\Ft + \frac{c^2}{3} \bnabla E &=& -\rho \kappaplus c \F
\label{eq:rad-*F}\\
{\frac{\partial p}{\partial t}} + (\u \cdot \bnabla) p - c_s^2
\left[{\frac{\partial \rho}{\partial t}} + (\u \cdot \bnabla)\rho\right]
&=& -(\gamma-1)\rho \kappa c(S-E)
\label{eq:entrop*y}
\end{eqnarray}
In Fig.\ 5 we show the eigenvalues from Fig.\ 3 with the corresponding ones
computed with the reduced equation set.

In fact, for the modes studied here, the terms $\partial_t E$ and
$\partial_t {\bf F}$ are also negligible.  Omitting these terms is
somewhat analogous to neglecting the displacement current in
hydromagnetics: we are giving up here the acoustic modes of the photon
fluid when we throw out those terms.  And, when we do discard those terms
in this reduced set of equations, we obtain the set of equations studied by
Asplund (1998), who finds instability only in the case of negative $g_*$.
Perhaps the absence of instability in situations like those we have
studied here comes from Asplund's use of a local approximation.

Instabilities of acoustic modes are found in
models of cooler stellar atmospheres with very similar descriptions to
the one given in this section.
Such studies have $\alpha=1$ and they typically
replace the description of the radiative terms by the diffusion limit
or the optically thin limit, in which Newton's law of cooling may be used.
We believe that the instabilities we see at small horizontal wavenumber
are descendants of those thermoacoustic instabilities (Umurhan, 1998)
with possibly an admixture of some form of $\kappa$-mechanism (Unno
{\it et al.} 1979).

It is not clear why the treatment of Marzek does not show at least these
instabilities inherited from cooler conditions.  It may be that this
is a matter of underlying structures.  The cool-star studies are largely
confined to polytropic layers and we have seen that the full atmosphere
typically has a composite structure.  On the other hand, we have run
calculations in layers confined to $z>0$, that is with the polytropic
part cut out, and we still find instabilities.   The fact that these
were not seen (or at least not reported by) Marzek is therefore
disquieting.  He did work with the transfer equation and so his results
contain the influence of radiative viscosity that we have neglected. On
the other hand, at large horizontal wavenumbers, when the modes become
optically thin, these effects should not be large.  This discrepancy
remains unresolved.

These results show why we felt that it is appropriate to reopen the
stability discussion here.  We claim that hot stellar atmospheres are
subject to instabilities in the form of growing waves.  Another
contributor to this volume, N. Shaviv, has also offered this general
conclusion, though as yet we have not made detailed comparisons with this
work.  We have here confined our calculations to compact domains, thus
finding the absolute instabilities, but we suspect that there are also
strong drift instabilities.  These are harder to study by numerical
means, but we are now undertaking some analytic work on that issue.

If it is agreed that radiative instabilities do occur in the atmospheres
of the hottest stars, it is of interest to look ahead to the possible
nonlinear developments one may expect from such instabilities.  Cassinelli
(1985) has argued that hot stars have spots on the grounds that they
exhibit the phenomena that go with them in the solar case, such as
coronas and other high excitation features.  Since one does not expect the
conditions for ordinary convection in hot stellar atmospheres, another
mechanism is needed if this suggestion is correct.

The acoustic instabilities we have discussed here are likely to produce
strong density inhomogeneities with perhaps the kind of bubbling that one
sees in fluidized beds (Prendergast and Spiegel, 1973).  Since hot stars
are typically rapid rotators, we may expect the usual symbiosis of
vigorous stellar fluid dynamics and stellar rotation that is believed to
produce dynamo action. Cassinelli's conjecture seems to us to have a
strong chance to be correct.  We believe that photofluiddynamic
instabilities are likely to be of interest in the study of hot stars and
disks.  While the need for further stability calculations remains,
we feel that numerical simulation in the nonlinear regime may now be
a fruitful pursuit.  We hope to have results on such calculations with
photon bubbles and radiative vortices (phortices) in time for Giora's
hundred and twentieth birthday.

L.T. acknowledges support from an NSF postdoctoral fellowship.  We
were participants in the Geophysical Fluid Dynamics summer school
at Woods Hole Oceanographic Institution during the completion of
this work.

\section*{References}

\n
Anderson, J.L. and Spiegel, E.A. 1972, ApJ 171, 127.

\n
Arons, J. 1992, ApJ 388, 561.


\n
Asplund, M. 1998, AA 330, 641.

\n Cassinelli, J.P. 1985 in {\it The Origin of Nonradiative
Heating/Momentum in Hot Stars}, A.B. Underhill and A.G. Michalitsianos,
eds.\ (NASA 2358) 2.

\n
Cayrel, R. and Steinberg, M., eds. 1976 {\sl Physique des
Mouvements dans les Atmospheres Stellaires}, Colloques Internationaux du
C.N.R.S., No. 250; (see therein the papers of Castor, Hearn and Spiegel).


\n
Davidson, J.F. \& Harrison, D. 1963 {\sl Fluidized Particles,}
(Cambridge University Press)

\n
Hsieh, S.-H. Thesis. Columbia Astronomy.

\n
Hsieh, S.-H. \& Spiegel, E.A. 1976, ApJ 207, 244.

\n
Huang, S.-S. \& Struve, O. 1960 in {\sl Stellar Atmospheres},
J.L. Greenstein, ed.\ (University of Chicago) 300.

\n
Humphreys, W.J. {\sl Physics of the Air} (Dover Publications, 1964),
pp. 51, ff.

\n
Lamb, H. 1932 {\sl Hydrodynamics} (6th edition) (Dover).

\n
Marzek, C. 1977 Thesis. Columbia Univeristy. Department of Astronomy.

\n
Moore, D.W. \& Spiegel, E.A. 1964, ApJ 139, 48.

\n
Poyet, J.-P. and Spiegel, E.A. 1979, Astron. J. 84, 1918.

\n
Prendergast, K.H. and Spiegel, E.A. 1973, Comm.\ Ap.\ Space Phys.\ 5, 43.

\n
Simon, R. 1963, J.\ Quant.\ Spectr.\ Rad.\ Transf.\  3, 14.

\n
Spiegel, E.A. 1964, ApJ 139, 959.

\n
Spiegel, E.A., ``Photoconvection,'' in {\sl Problems in Stellar
Convection}, E.A.  Spiegel and J.-P. Zahn, eds., Springer-Verlag,
Berlin `Heidelberg' New York, 1977.

\n
Umurhan, O.M. 1998, Thesis, Columbia University, Astronomy Department.


\n
Underhill, A.B. 1959 MNRAS 109, 563.

\n
Unno, W., Osaki, Y., Ando, H \& Shibahashi, H. 1979 {\sl Nonradial
Oscillations of Stars} (Univ. of Tokyo Press).

\newpage

\section*{Figure Captions}
\n{Figure 1. Dispersion relation of the fundamental acoustic mode for
$\alpha = 0.7$.  The top panel shows the acoustic frequency and
the bottom panel is the instability growth rate.  This is the case
of the polytrope atmosphere.}

\n{Figure 2. Dispersion relation of the fundamental acoustic mode for
$\alpha = 0.7$.  The top panel shows the acoustic frequency and
the bottom panel is the instability growth rate. This is the case
of the isothermal atmosphere.}

\n{Figure 3. Dispersion relation of the fundamental acoustic mode for
$\alpha = 0.7$.  The top panel shows the acoustic frequency and
the bottom panel is the instability growth rate. This is the case
of the composite atmosphere, where
the transition from the polytrope to the isothermal atmosphere occurs
at $z = 0$.}

\n{Figure 4.  Eigenfunctions of the fundamental acoustic mode for
$\alpha = 0.7$.  The left and right panels correspond to real and
imaginary parts.  This is the case of the composite atmosphere, where
the transition from the polytrope to the isothermal atmosphere occurs
at $z = 0$.}

\n{Figure 5. Comparison of dispersion relations between the full
equations and the reduced equations.  Solid lines are
the full calculation and the dashed lines are the reduced one.
This is for the composite atmosphere at $\alpha = 0.7$}

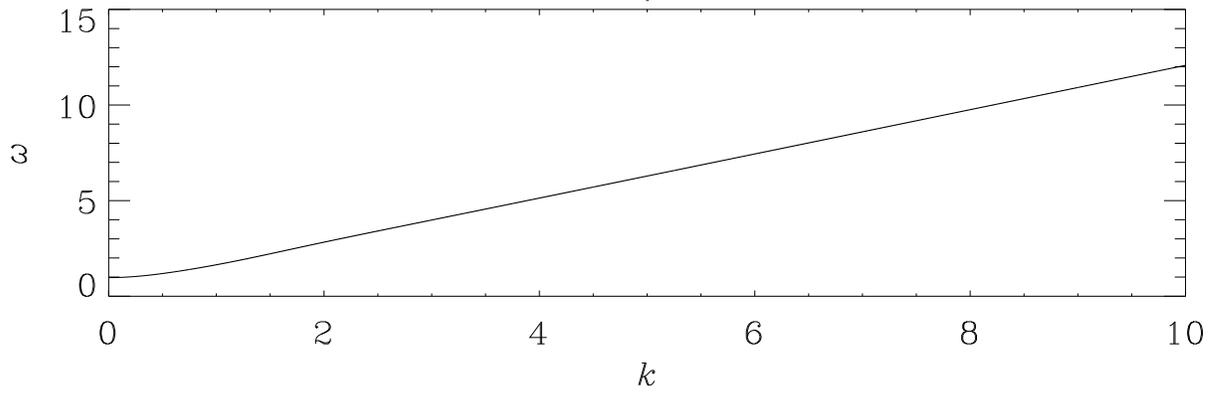
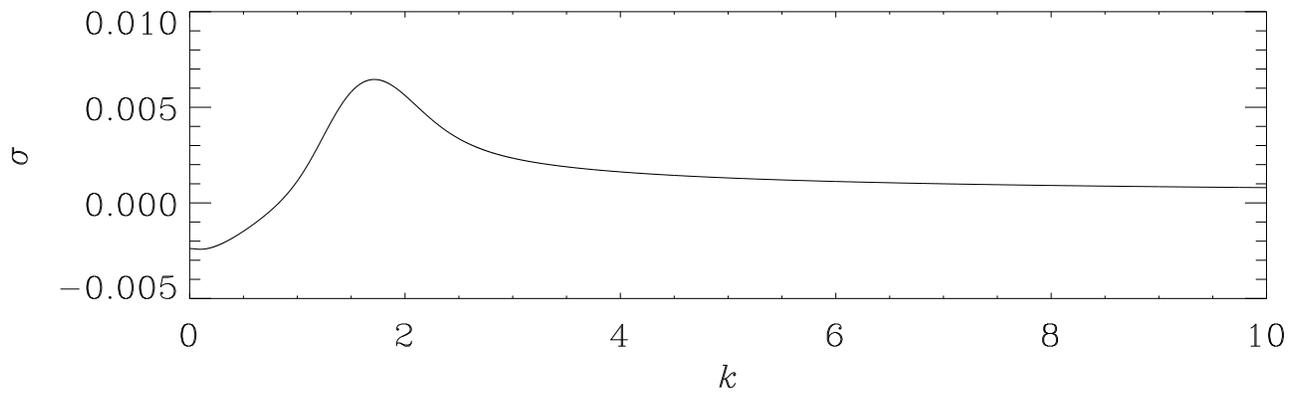
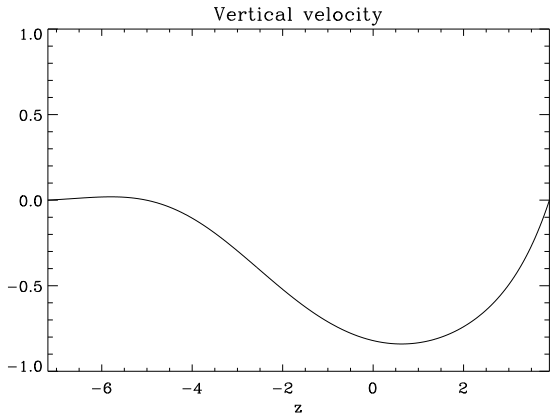
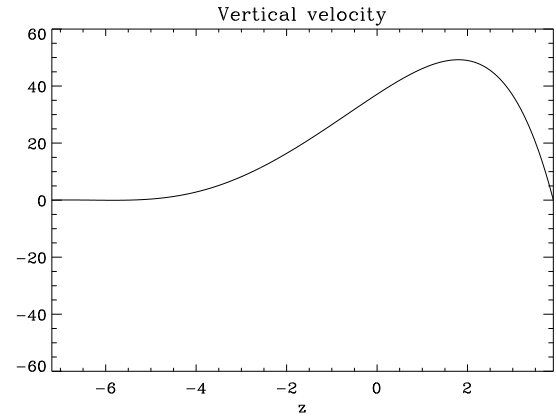
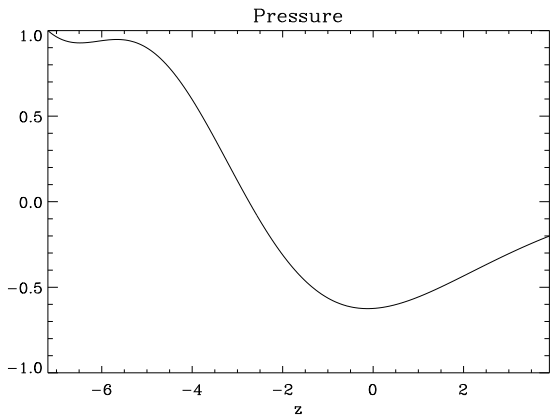
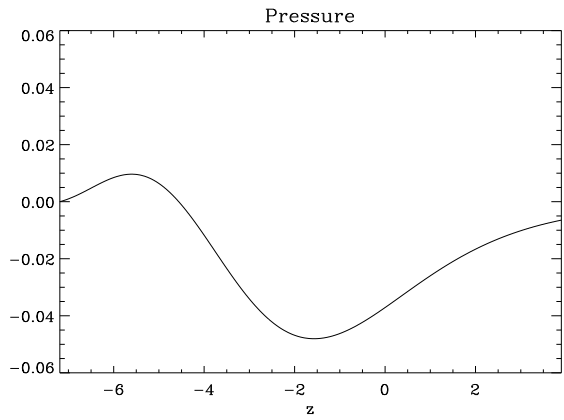
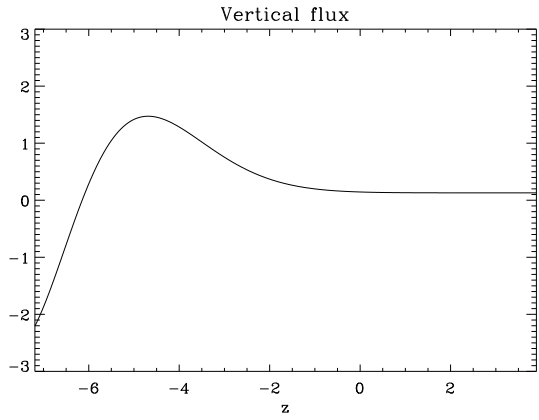
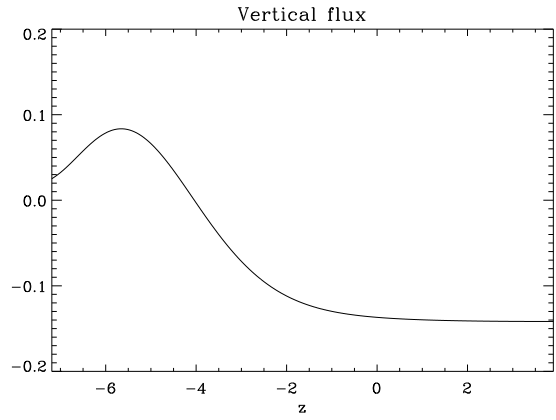
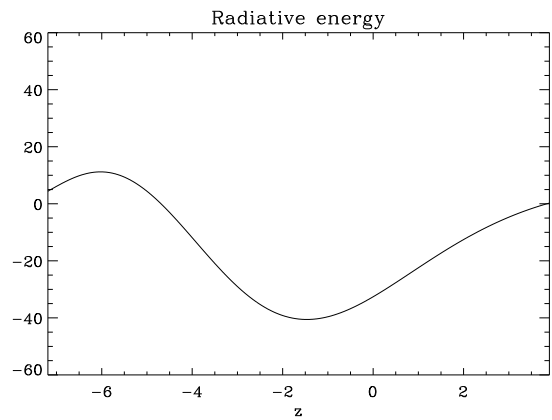
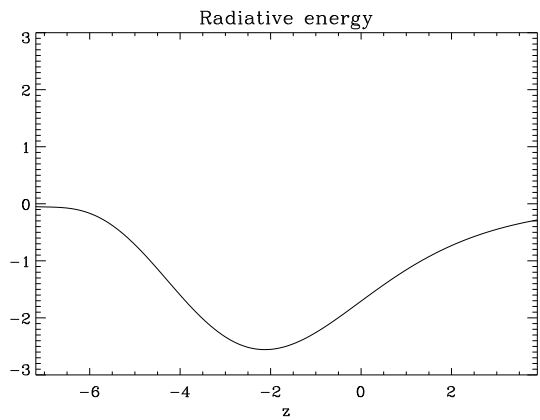
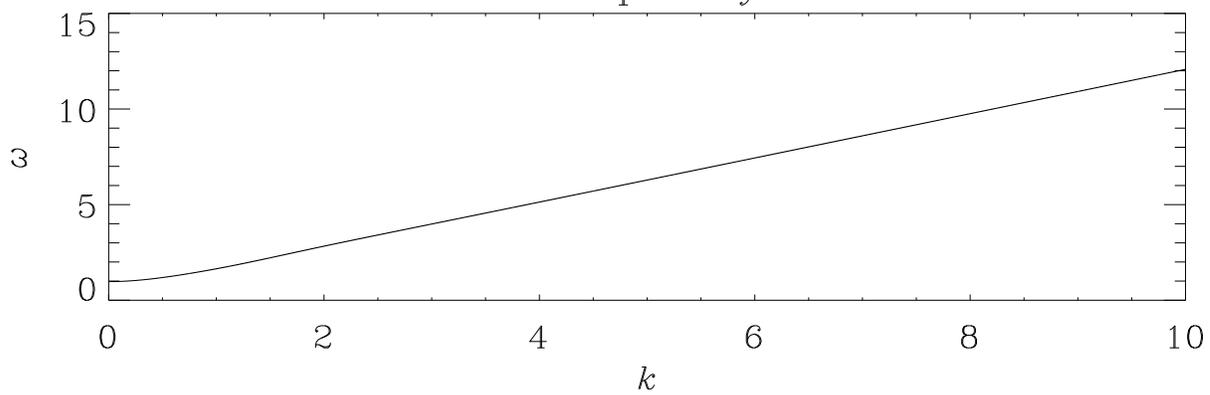
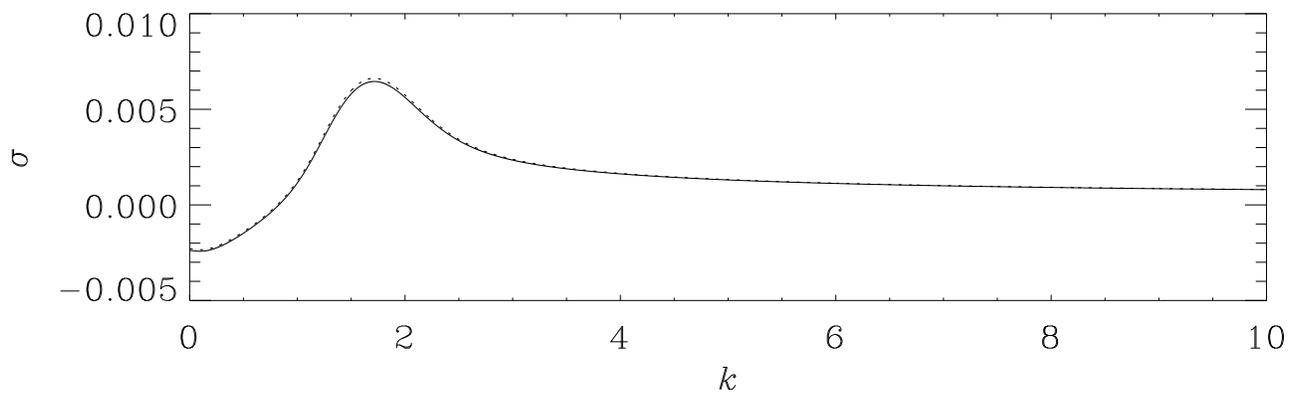
\end{document}